# Lindemann's Rule Applied to the Melting of Crystals and Ultra-Stable Glasses

Robert F. Tournier[1,2]

[1]Univ. Grenoble Alpes, Inst. NEEL, F-38042 Grenoble Cedex 9, France

[2]CNRS, Inst. NEEL, F-38042 Grenoble, France

E-mail address: robert.tournier@creta.cnrs.fr

**Abstract**: The ratio of the mean square amplitude root of thermal vibrations and the interatomic distance is a universal constant $\delta_{ls}$ at the melting temperature $T_m$. The classical Gibbs free energy change completed by a volume energy saving $\varepsilon_{ls}$ (or $\Delta\varepsilon_{lg}$)$\times\Delta H_m$ that governs the liquid to solid and liquid to ultra-stable glass transformations leads to a universal constant equal to $\delta_{ls}$ (or $\delta_{lg}$), $\Delta H_m$ being the crystal melting enthalpy. The minimum values 0.217 of $\varepsilon_{ls}$ and 0.103 of $\delta_{ls}$ are used to predict ultra-stable glass formation in pure metallic liquid elements at a universal reduced temperature $\theta_g = (T_g - T_m)/T_m = -0.6223$.

1- **Introduction**

The dependence of the liquid supercooling temperature on the superheating rate shows the existence of long-lived metastable nuclei surviving above the melting temperature $T_m$ [1]. The classical Gibbs free energy change cannot predict the presence of such entities without introducing a complementary negative contribution $v\times\Delta p$ varying with $\theta^2 = (T-T_m)^2/T_m^2$, v being the nucleus volume and $\Delta p$ a complementary Laplace pressure [2,3]. Crystallisation and melting are initiated by the formation of solid or liquid growth nuclei accompanied by a volume change that is expected to obey to Lindemann's rule [4]. Lindemann's description shows that the ratio of the mean square amplitude root of thermal vibrations and the interatomic distance is a universal constant $\delta_{ls}$ at the melting temperature $T_m$.

The critical complement $v\times\Delta p$ associated with crystallisation at $T = T_m$ has been determined for many pure liquid elements and glass-forming melts as being equal to $v\times\varepsilon_{ls0}\times\Delta H_m/V_m$ with $\varepsilon_{ls0}$ being a numerical critical fraction of the melting heat $\Delta H_m$. The coefficient $\varepsilon_{ls0} = 0.217$ is the same for many liquid elements [2], while it is much larger than 1 and smaller than 2 in many glass-forming melts, as shown for 84 examples Tables 2 and 3 in [5]. The objective of this study is to relate $\varepsilon_{ls0}$ to the Lindemann ratio $\delta_{ls}$.

The ultra-stable glass state is described as a thermodynamic equilibrium between crystal and liquid states, which would be attained by supercluster formation and their percolation after a very long annealing time at the Kauzmann temperature $T_K$ [5], or by quenching the melt from above $T_m$ and annealing it at the formation temperature $T_{sg}$ of this phase [6]. The optimum formation temperature $T_{sg}$ leading to the higher density is always equal to the Kauzmann temperature of strong glasses, and very often to that of fragile glasses. The enthalpy that is recovered at the glass transition temperature $T_g$ is equal to $(\varepsilon_{ls0}-\varepsilon_{gs0})\times\Delta H_m$ in strong glasses and to $1.5\times(\varepsilon_{ls0}-\varepsilon_{gs0})\times\Delta H_m$ in fragile glasses with $\varepsilon_{gs0}$ being the critical fraction of melting heat leading to crystallisation of a virtual glass at $T_m$. This description agrees, in principle, with the scheme of a random first-order phase transition hidden below $T_g$ occurring at the Kauzmann temperature $T_K$ viewed as the true glass transition at equilibrium [7–9].

Various microscopic models prove the existence of a phase transition at $T_g$ [10–16]. Liquid-glass transformation is treated as a percolation-type phase transition with the formation of dynamic fractal structures near the percolation threshold. We do not examine the problem of percolation of liquid entities, because the total volume change of all of them, depending on their nucleation rate and their atom number n, is limited by the effective volume change available at the glass transition. A criterion analogous to the Lindeman criterion of melting was proposed by Sanditov for the softening transition from the glass equilibrium to the liquid state [17]. We are also able to determine Lindemann's constant from our model and show whether it agrees with this model of glass melting.

The universal constant $\delta_{ls} = 0.103$ obtained for pure metallic elements at their melting temperatures $T_m$ is used to build a model for their vitrification. The Gibbs free energy change below $T_g$ cannot include any variation of structural relaxation enthalpy, because $\delta_{ls}$ cannot be lower than this minimum value.

## 2- The application of Lindemann's rule

The Gibbs free energy change for the formation of a condensed supercluster giving rise by growth to a crystal from a liquid droplet of radius R is defined by Eq. (1):

$$\Delta G_{2ls}(R,\theta) = \frac{\Delta H_m}{V_m}(\theta - \varepsilon_{nm})4\pi\frac{R^3}{3} + 4\pi R^2 \frac{\Delta H_m}{V_m}(1+\varepsilon_{nm})(\frac{k_B V_m \ln K_{ls}}{36\pi \times \Delta S_m})^{1/3} \quad (1)$$

where $\Delta H_m$ is the melting heat, $V_m$ the molar volume, $\theta = (T-T_m)/T_m$ the reduced temperature, $\varepsilon_{nm}$ the fraction of the melting enthalpy associated with a spherical supercluster of radius R containing n atoms, $k_B$ the Boltzmann constant, $\Delta S_m$ the melting entropy and $\ln(K_{ls}) \cong 90 \pm 2$ in metallic liquid elements [2,18].

The thermal variation of $\varepsilon_{nm}$ is given by Eq. (2):

$$\varepsilon_{nm}(\theta) = \varepsilon_{nm0}(1 - \frac{\theta^2}{\theta_{0m}^2}) \quad (2)$$

where $\varepsilon_{nm0}$ obeys Eq. (3) for n > 147, $\varepsilon_{nm0} = \varepsilon_{ls0}$ being the critical value, $n_c$ the critical atom number given by Eq. (4):

$$\varepsilon_{nm0} = \varepsilon_{ls0} \times (\frac{n_c}{n})^{1/3} \quad (3)$$

$$n_c = \frac{8(1+\varepsilon_{ls})^3}{27(\theta - \varepsilon_{ls})^3} \frac{N_A k_B \ln K_{ls}}{\Delta S_m} \quad (4)$$

where $\Delta S_m$ is the melting entropy of crystals and $\varepsilon_{ls}$ the critical enthalpy saving coefficient, given as a function of the reduced temperature by Eq. (5) [2]:

$$\varepsilon_{ls}(\theta) = \varepsilon_{ls0}(1 - \frac{\theta^2}{\theta_{0m}^2}) \tag{5}$$

Eq. (1) is written as a function of the atom number n instead of the supercluster radius, with $N_A$ being the Avogadro number [5]:

$$\Delta G_{nls}(n,\theta,\varepsilon_{nls}) = \Delta H_m \frac{n}{N_A}(\theta - \varepsilon_{nls}) + \frac{(4\pi)^{1/3}}{N_A}\Delta H_m(1+\varepsilon_{nls})\left[\frac{N_A k_B \ln(K_{lg})}{36\pi\Delta S_m}\right]^{1/3}(3n)^{2/3} \tag{6}$$

The entropy of fusion given by $-(d\Delta G_{nls}/dT)_p$ is equal to $n\times\Delta S_m/N_A$ because $d\varepsilon_{nls}/dT$ is equal to zero for $\theta = 0$, and then the surface energy is constant during the solid-liquid transformation. The melting heat of supercluster surface atoms is not weakened. Superclusters are viewed as superatoms. The free electrons in a superatom occupy orbitals that are defined by the entire group of atoms of the supercluster, rather than by each individual atom separately [19,20]. We consider that the melting heat is proportional to the number n of atoms forming a superatom. These equations have been successfully applied to the crystallisation of pure metallic undercooled liquids. Such nuclei containing magic atom numbers govern the undercooling rate [3].

Lindenmann's rule predicts that the supercluster radius R is increased by the root of the mean square amplitude of atom vibrations when it is melted. An increase $\Delta R$ of the radius R is applied to the surface energy given by Eq. (1). Eq. (7) is obtained assuming that the surface energy does not vary during the transformation from solid to liquid:

$$1 + \varepsilon_{nm0} = (1 + \frac{\Delta R}{R})^2$$

$$\delta_{ls} = \frac{\Delta R}{R} = (1 + \varepsilon_{ls0})^{1/2} - 1 \tag{7}$$

The coefficient $\varepsilon_{nm0}$ is always larger than its critical value $\varepsilon_{ls0}$ at $T_m$ when $n < n_c$. Then, the corresponding Lindemann ratio is larger than its critical value. For the weakest values of $\varepsilon_{ls0}$, we have $\varepsilon_{ls0} \cong 2\times\Delta R/R$.

The critical enthalpy saving associated with the Laplace pressure change accompanying a supercluster condensation having the critical radius for crystal growth is given below $T_g$ by Eq. (8):

$$\varepsilon_{gs}(\theta)\frac{\Delta H_m}{V_m} = \varepsilon_{gs0}(1 - \frac{\theta^2}{\theta_{0g}^2})\frac{\Delta H_m}{V_m} \tag{8}$$

The enthalpy change at $T_g$ transforms Eq. (2) into Eq. (8). The indices ls and gs are related to the undercooled liquid crystallisation and to the glass crystallisation, respectively. The numerical coefficient $\varepsilon_{ls0}$ in Eq. (2) above $T_g$ becomes weaker below $T_g$ and is transformed into $\varepsilon_{gs0}$ while the temperature $T_{0m}$ deduced from $\theta_{0m} = (T_{0m} - T_m)/T_m$ is reduced and is equal to $T_{0g}$ in $\theta_{0g}$. These enthalpy saving coefficients $\varepsilon_{ls}(\theta)$ and $\varepsilon_{gs}(\theta)$ are equal to zero at $T \leq T_{0m}$ and $T \leq T_{0g}$ and to $\varepsilon_{ls0}$ and $\varepsilon_{gs0}$ respectively at the crystal melting temperature $T_m$. The enthalpy change, with the index lg, is associated with the ultra-stable glass-to-liquid transformation and is equal to the difference between Eqs. (2) and (8) given by Eq. (9) [5,6]:

$$\Delta \varepsilon_{lg}(\theta) \times \Delta H_m = (\varepsilon_{ls} - \varepsilon_{gs}) \times \Delta H_m = \varepsilon_{ls0} \times \Delta H_m (1 - \frac{\theta^2}{\theta_{0m}^2}) - \varepsilon_{gs0} \times \Delta H_m (1 - \frac{\theta^2}{\theta_{0g}^2}) \quad (9)$$

The endothermic latent heat $\Delta \varepsilon_{irr}$ recovered at $T_g$ during the transition from ultra-stable glass state to liquid is the maximum value $(\varepsilon_{ls0} - \varepsilon_{gs0}) \times \Delta H_m$ of $\Delta \varepsilon_{lg}$ at $T_m$. It is enhanced by the enthalpy available between the temperature where $\Delta \varepsilon_{lg}(\theta)$ in Eq. (9) is equal to zero and $T_g$ which is equal to $0.5 \times (\varepsilon_{ls0} - \varepsilon_{gs0}) \times \Delta H_m$ for fragile glasses and to zero for strong glasses. The Gibbs free energy change is given at $T_g$ by Eq. (10) before the transformation of the ultra-stable glass into a liquid state [5]:

$$\Delta G_{n\lg}(n, \theta, \Delta \varepsilon_{n\lg}) = \Delta H_m \frac{n}{N_A}(-\Delta \varepsilon_{irr}) + \frac{(4\pi)^{1/3}}{N_A} \Delta H_m (1 + \Delta \varepsilon_{irr}) \left[ \frac{N_A k_B \ln(K_{\lg})}{36\pi \Delta S_m} \right] 3n)^{2/3} \quad (10)$$

Eq. (7) can be applied to the melting of ultra-stable glasses replacing $\varepsilon_{nm0}$ by the critical coefficient $\varepsilon_{ls0}$ for crystal melting, by $\Delta \varepsilon_{irr} = (\varepsilon_{ls0} - \varepsilon_{gs0})$ for devitrification of ultra-stable strong glasses and by $\Delta \varepsilon_{irr} = 1.5 \times (\varepsilon_{ls0} - \varepsilon_{gs0})$ for devitrification of fragile glasses [6]. These last coefficients define the fraction $\Delta \varepsilon_{irr}$ of $\Delta H_m$ associated with the endothermic latent heat accompanying these transformations at $T_g$. A softening transition occurs at $T_g$ and the Lindemann constant $\delta_{lg}$ of the liquid-glass transformation depends on the difference of Lindemann's constants $\delta_{ls}$ and $\delta_{gs}$ of two liquid states of the same substance above and below $T_g$. Eqs. (11) and (12) are respectively followed by strong and fragile glass-forming melts:

$$\delta_{lg} = \delta_{ls} - \delta_{gs} \quad (11)$$

$$\delta_{lg} = 1.5 \times (\delta_{ls} - \delta_{gs}) \quad (12)$$

### 3- Typical values of the Lindemann constant for crystal and ultrastable glass melting

*3.1 The Lindemann constant of pure liquid elements*

The energy saving coefficient $\varepsilon_{ls0}$ was determined in 2007 as being constant and equal to 0.217 for about 30 pure metallic liquid elements [2]. None explanation was given. Applying Eq. (7), $\varepsilon_{ls0}$ corresponds to a Lindemann constant equal to 0.103, which is in good agreement with other determinations [21–23]. Then, the enthalpy saving coefficient cannot be smaller than 0.217. This property is used to propose, in Section 4, that a new family of ultra-stable glasses composed of all pure liquid elements exists.

### *3.2 The Lindemann constant of ultra-stable fragile glasses at $T_g$*

The endothermic enthalpy coefficients of ultra-stable glasses have been determined for many fragile glass-forming melts. They are equal to $1.5\times(\varepsilon_{ls0}-\varepsilon_{gs0})$ and known for 49 non-metallic and 23 metallic glass-forming melts. The coefficients $\varepsilon_{ls0}$ and $\varepsilon_{gs0}$ are listed in Tables 2 and 3 in [5]. The corresponding Lindemann constants $\delta_{lg}$ given by Eqs. (7) and (12) are plotted versus the reduced glass transition temperature $\theta_g$ in Figure 1. The values calculated by Sanditov vary between 0.11–0.14 [17]. Our results cover these values.

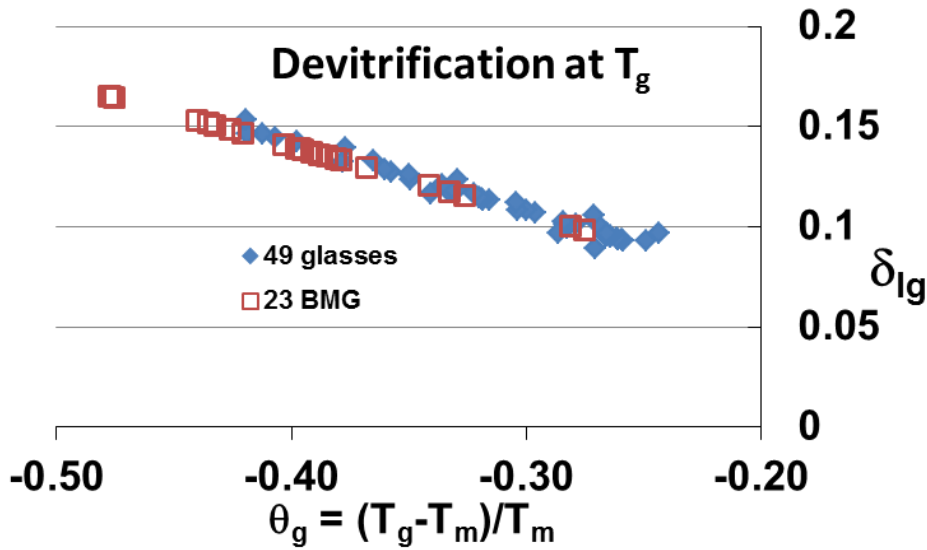

**Figure 1:** *The Lindemann constant $\delta_{lg}$ is calculated with Eqs. (7) and (12). The coefficient $\varepsilon_{nls0}$ in Eq. (7) has been replaced by $\Delta\varepsilon_{irr}= 1.5\times(\varepsilon_{ls0}-\varepsilon_{gs0})$. The $\delta_{lg}$ of bulk metallic and non-metallic glasses are equal for the same reduced value of $T_g$.*

### *3.3 The Lindemann constant of fragile glass-forming melts at their melting temperature $T_m$*

The enthalpy saving coefficient $\varepsilon_{ls0}$ is used to calculate $\delta_{ls}$ at the melting temperature of glass-forming melts. Many $\delta_{ls}$ follow a linear law as a function of $\theta_g$ in Figure 2: $\delta_{ls} = 0.307\times\theta_g+0.735$ because of the existence of a scaling law for a large fraction of glass-forming melts [5]. The points outside the straight line correspond to those having a latent heat at $T =T_g$.

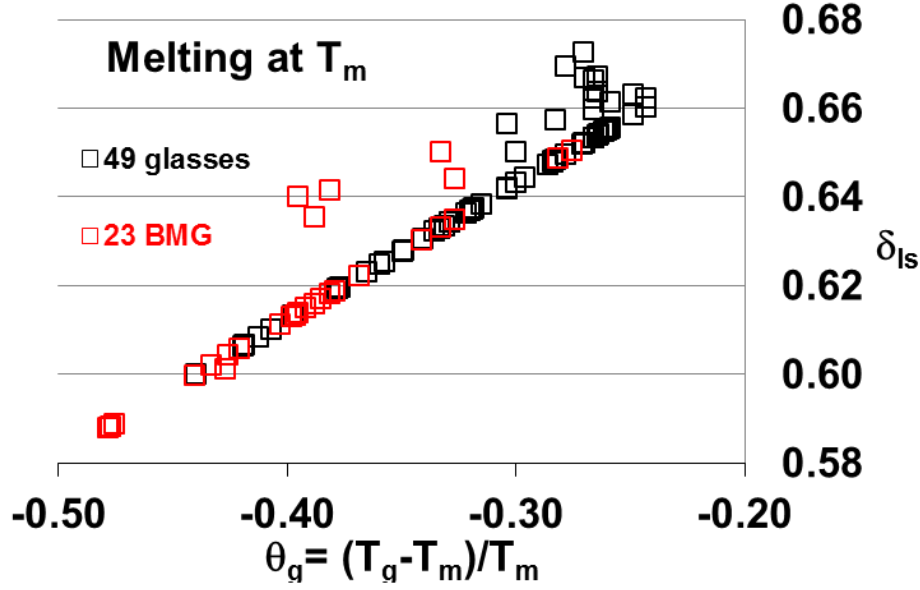

**Figure 2**: *The Lindemann constant $\delta_{ls}$ calculated with Eq. (7) at the melting temperature $T_m$ of glass-forming melts, plotted as a function of the reduced glass transition temperature $\theta_g$. All points on the straight line follow the scaling law $\varepsilon_{ls0}$-$\varepsilon_{gs0}$ =0.5×$\theta_g$. The other glasses have a small exothermic latent heat during cooling. The value of $\delta_{ls}$ depends on this enthalpy excess coefficient.*

These values of $\delta_{ls}$ are about 6 times larger than the minimum value and increase with $\theta_g$. The vibration amplitude attains 66% of the characteristic interatomic distance for the largest ratio $T_g/T_m$. We know from Inoue's work that alloys having the highest $T_g/T_m$ have three features in their alloy components, i.e. multicomponent systems, significant atomic size ratios above 12% and negative heats of mixing [24,25]. These high values of $\delta_{ls}$ could be due to a strong coupling of pairs of different atoms in the liquid, which does not exist in pure liquid elements.

### *3.4 The Lindemann constant of strong glass-forming melts at $T_g$*

The endothermic enthalpy coefficient of a strong ultra-stable glass at $T_g$ is equal to ($\varepsilon_{ls0}$-$\varepsilon_{gs0}$), while the specific heat jump at $T_g$ is equal to $2\times\Delta S_m\times(\varepsilon_{ls0}$-$\varepsilon_{gs0})/\theta_g$. The Lindemann constant $\delta_{lg}$ is then proportional to the specific heat jump. This jump is known to be much smaller than that of a fragile glass. It is important to verify whether these glasses obey Eq. (7), and the values of $\delta_{lg}$ can be smaller than the minimum value. Nine strong glasses have been studied and some values of ($\varepsilon_{ls0}$-$\varepsilon_{gs0}$) are reported in Table 1 in [5]. Five of them ($CaAl_2Si_2O_8$, $As_2Te_{3.13}$, $CaMgSi_2O_6$, $Zr_{46.75}Ti_{8.25}Cu_{7.5}Ni_{10}Be_{27.5}$, $Au_{77}Ge_{13.6}Si_{9.4}$) have $0.099 < \delta_{lg} < 0.13$; and four of them ($SiO_2$, $BeF_2$, $NaAlSi_3O_8$, $GeO_2$) have $0.029 < \delta_{lg} < 0.085$. The endothermic latent heat at $T_g$ corresponds to a softening transition and the $\delta_{lg}$ of the glass state is the difference of Lindemann's constants of two liquid states of the same substance, above and below $T_g$ as shown by Eq. (11). Sanditov's proposal is verified.

### 4- Application of Lindemann's rule to the vitrification of pure liquid elements

Recent work renews earlier findings of glass formation in pure metals of small size and thickness [26–36]. There is a need for a fundamental understanding of the resistance to crystallisation of these glasses [37]. The glass transition temperature $T_g$ is unknown. Our model needs in principle to know $T_g$ in order to be applied. Lindemann's rule is used to determine this temperature. The glass transition transforms $\varepsilon_{ls0}$ in Eq. (5) to $\varepsilon_{gs0}$ in Eq. (8). In pure liquid elements, Lindemann's rule shows that this change is not possible. Eq. (8) can be used below $T_g$ with $\varepsilon_{gs0} = \varepsilon_{ls0} = 0.217$. There is no structural relaxation enthalpy because ($\varepsilon_{ls0} - \varepsilon_{gs0}$) is equal to zero [5,6]. The liquid elements are strong with $\theta_{0m} = -2/3$ ($T_{0m} = T_m/3$) and become stronger glasses below $\theta_g$. The reduced temperature $\theta_{0m} = -2/3$ in Eq. (5) is changed in $\theta_{0g} = -1$ in Eq. (8) as it occurs in many strong glasses because the relaxation time follows an Arrhenius law with a Vogel-Fulcher-Tammann temperature equal to zero below $T_g$.

All glasses obey Eq. (13) obtained by combining Eq. (5) with the homogeneous nucleation temperature $\theta = [\varepsilon_{gs}(\theta)-1]/3$ deduced from Eq. (1) or Eq. (8). The minimum value -2/3 of $\theta_{0m}$ ($T_{0m} = T_m/3$) in Eq. (13) is chosen for liquid elements. It leads to $\theta = \theta_1 = -2/3$. The reduced glass transition temperature $\theta_g = \theta_2$ is also given by Eq. (13) with $\theta_{0g} = -1$ [5,6]:

$$\varepsilon_{ls0} = \frac{3 \times \theta_1 + 2}{1 - \theta^2/\theta_{0m}^2} \qquad \varepsilon_{gs0} = \frac{3 \times \theta_2 + 2}{1 - \theta^2/\theta_{0g}^2} \qquad (13)$$

The variation of $\varepsilon_{gs0}$ with $\theta_g$ is plotted in Figure 3 for $\theta_{0g} = -1$. The value of $\theta_g$ corresponding to $\varepsilon_{gs0} = 0.217$ is $-0.6223$. This could be a universal value of the reduced glass transition temperature of pure liquid elements corresponding to $T_g = 0.3777 \times T_m$ because the condition $\theta_{0g} = -1$ is always respected by strong glass-forming melts respecting $\theta_{0m} \leq -2/3$. The reduced temperature $\theta_{0m}$ in Eq. (13) is equal to $-2/3$ corresponding to the value $T_m/3$ of the Vogel-Fulcher-Tammann (VFT) temperature. The predicted glass transition temperature of liquid elements is weak as compared to that of classical strong glasses represented in Fig. (2). Their VFT temperature $\theta_{0m} = -2/3$ is still smaller than $T_g$. A much lower value of $T_g$ has been observed in colloidal suspensions with the VFT temperature equal to $T_g$ [38]. Eq. (13) applied for $\varepsilon_{gs0} = 0$ also leads to a glass transition temperature equal to the VFT temperature $T_m/3$ as shown in Figure (3). Even in this case, there is a link between glass transition and critical phenomena [10, 11, 38].

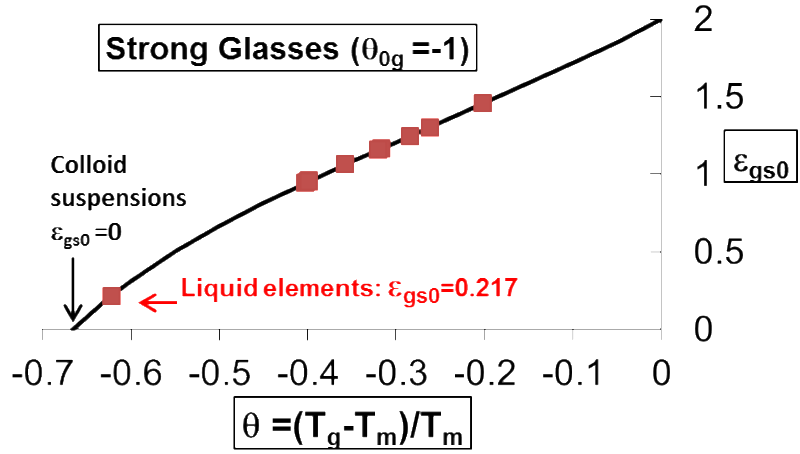

**Figure 3**: *Eq. (13) is used to represent $\varepsilon_{gs0}$ as a function of the reduced glass transition temperature $\theta_g$ for $\theta_{0g} = -1$. The point corresponding to the universal value 0.217 of $\varepsilon_{gs0}$ corresponds to $\theta_g = -0.6223$. The other points belong to classical strong glasses. The lowest glass transition temperature $\theta_g = -2/3$ for $\varepsilon_{gs0} = 0$ has been observed in colloidal suspensions.*

Eq. (9) is transformed in Eqs. (14) and (15), which are universal equations describing the enthalpy change $\Delta\varepsilon_{lg}$ below $\theta_g$:

$$\Delta\varepsilon_{lg} = -1.25 \times \theta^2 \times 0.217 \quad \text{with } -2/3 < \theta < -0.6223 \tag{14}$$

$$\Delta\varepsilon_{lg} = 0.217 \times (1-\theta^2) \quad \text{with} -1 < \theta < -2/3 \tag{15}$$

A latent heat equal to $0.10506 \times \Delta H_m$ is predicted at the glass transition temperature. These equations are represented in Figure 4 together with Eqs. (2) and (5):

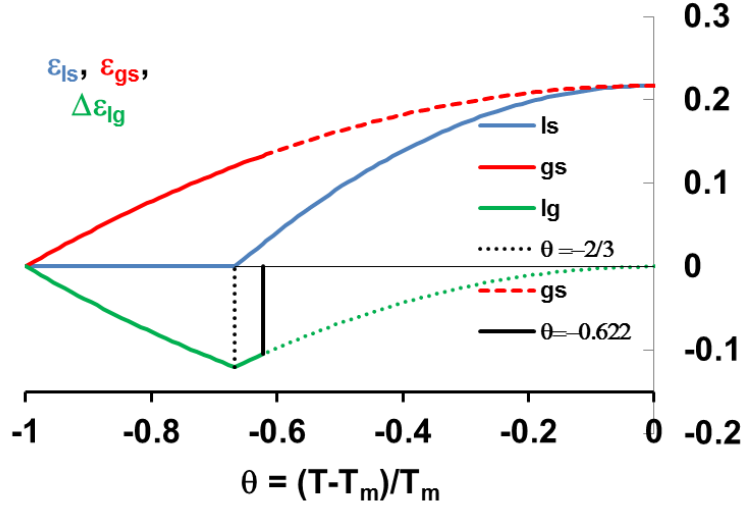

**Figure 4:** *The critical enthalpy saving coefficients $\varepsilon_{ls}$ and $\varepsilon_{gs}$ associated with crystal formation given by Eqs. (5) and (8) are plotted versus the reduced temperature between $\theta = -1$ and 0. The enthalpy saving coefficient $\Delta\varepsilon_{lg}$ associated with the ultra-stable glass state formation and given by Eq. (9) is also plotted versus $\theta$. At the glass transition, an exothermic heat is produced at equilibrium and probably reported at the final temperature after quenching.*

The Lindemann constant $\delta_{lg}$ at $T_g$ deduced from Eq. (7) is equal to 0.0512, because the latent heat is equal to $0.10506 \times \Delta H_m$.

Eq. (14) is nearly a straight line in Fig. (4) below $\theta = -2/3$ which can be approximated by a linear function of $\theta$ in Eq. (16):

$$\Delta\varepsilon_{lg} = 0.36165 \times (1+\theta) \qquad (16)$$

The quenching time is so weak that the exothermic latent heat could be delivered at the end of quenching process. The $\Delta\varepsilon_{lg}$ cannot vary with $\theta^2$ at low temperatures because the entropy associated with the latent heat calculated at $\theta_g = -0.6223$ has to stay equal to $0.10506 \times \Delta H_m/(1-0.6223)/T_m = 0.278 \times \Delta H_m/T_m$ at temperatures lower than $T_m/3$ with a fixed Kauzmann temperature. This condition is respected by Eq. (16). The enthalpy jump at room temperature (T= 300 K) is represented for several liquid metals in Figure 5.

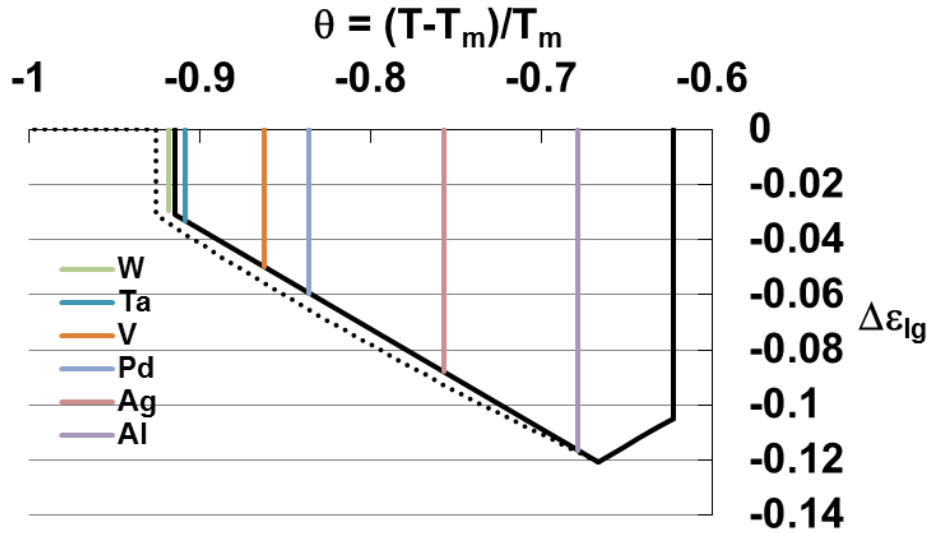

**Figure 5**: *The critical enthalpy saving coefficient $\Delta\varepsilon_{lg}$ plotted versus $\theta$. The reduced final temperature $\theta = (300-T_m)/T_m$ obtained after quenching is indicated together for the following liquid elements: W (-0.918), Ta (-0.909), V (-0.862), Pd (-0.836), Ag (-0.757) and Al (-0.678). The enthalpy saving coefficient $\Delta\varepsilon_{lg}$ calculated with Eq. (15) is represented by a dashed line. The reduced Kauzmann temperature is $\theta_K = -0.9143$.*

The latent heat per mole is $0.10506 \times \Delta H_m$. The enthalpy calculated from 0 K up to $\theta = -2/3$ is equal to $0.12055 \times \Delta H_m$ while that from $\theta = -2/3$ up to $\theta_g$ is equal to $0.01549 \times \Delta H_m$. $\Delta\varepsilon_{lg}$ has to be equal to zero below the Kauzmann temperature. The reduced Kauzmann temperature $\theta_K$ is then equal to $-0.9143$ and the frozen enthalpy from $T_K$ to $T_g$ is now equal to $0.10506 \times \Delta H_m$. The frozen entropy below $T_g$ and the exothermic latent heat occurring at $T_g$ in all vitrified liquid elements are equal to $0.278 \times \Delta S_m$, $\Delta S_m$ being the melting entropy of crystals. There is no structural relaxation enthalpy below $T_g$ and consequently no supplementary endothermic latent heat recovered at $T_g$ during warming.

The specific heat jump is equal to $\Delta H_m \times d\Delta\varepsilon_{lg}/dT$ and is defined by Eq. (17) [5]:

$$\Delta C_p = 2 \times 0.217 \times \frac{(T-T_m)}{T_m} \times (\frac{9}{4}-1) \times \Delta S_m \quad \text{in Joule/K/mole from } T_m/3 \text{ to } T_g,$$

$$\Delta C_p = 0.36165 * \Delta S_m \quad \text{in Joule/K/mole from 0 K to } T_m/3 \qquad (17)$$

These universal values are represented in Figure 6:

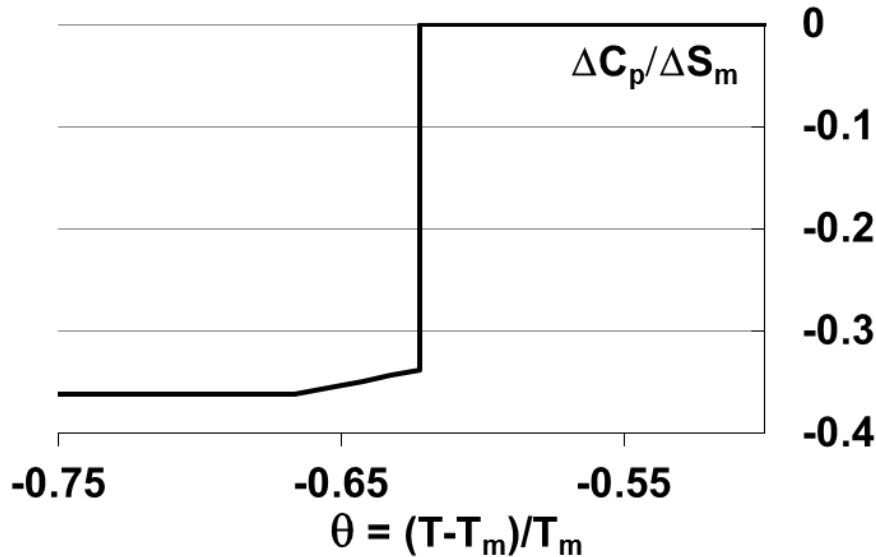

**Figure 6**: *The universal value $\Delta C_p/\Delta S_m$ of the reduced specific heat jump divided by the crystal melting entropy plotted versus $\theta = (T-T_m)/T_m$ for many liquid elements. The glass transition temperature of liquid elements is $0.3777 \times T_m$. The specific heat jump is maximum at $T= T_m/3$.*

**5- Conclusions**

The introduction, in the classical Gibbs free energy change, of an enthalpy saving associated with formation, in pure liquid elements, of solid superclusters, reveals the applicability of Lindemann's rule and leads to the expected theoretical value of the universal constant $\delta_{ls} = 0.103$. This new equation works because the melting heat of superclusters, acting as growth nuclei, and containing n atoms, is proportional to n, as expected in superatoms. The critical supercluster containing $n_c$ atoms melts at the temperature $T_m$ of crystals. The crystallisation could be induced by homogeneous or heterogeneous nucleation of superclusters containing magic atom numbers. Some of them survive above $T_m$ when their enthalpy saving coefficient varying as $(n_c/n)^{1/3}$ is much larger than that of the critical supercluster containing $n_c$ atoms, and when the liquid superheating rate is too small.

Lindemann's rule is used to predict the glass transition temperature $T_g= 0.377 \times T_m$ of liquid elements and universal thermodynamic properties of these glasses. These glasses are ultra-stable because they have no structural relaxation enthalpy.

Lindemann's constants $\delta_{ls}$ of glass-forming melts at $T_m$ are much higher than 0.103 when the ratio $T_g/T_m$ increases. This information could be the sign of strong pairing of atoms of different nature.

The devitrification of ultra-stable glasses at $T_g$ is associated with Lindemann's constant $\delta_{lg}$ that is equal in strong glasses to the difference of Lindemann's constants $\delta_{ls}$ and $\delta_{gs}$ of two liquid states of the same substance above and below $T_g$. This finding is in agreement with Sanditov's recent work, which considers that the softening transition at $T_g$ is somewhat similar to melting.